\documentclass[aps,pra,showpacs,superscriptaddress,preprint]{revtex4}
\usepackage{amsmath}
\usepackage{epsf}
\usepackage{epsfig}

\catcode`ð=\active
 \defð{\u{g}}
 \catcode`Ð=\active
 \defÐ{\u{G}}
 \catcode`Ý=\active
 \defÝ{\. I}
 \catcode`ö=\active
 \defö{\"{o}}
 \catcode`Ö=\active
 \defÖ{\"O}
 \catcode`ü=\active
 \defü{\"{u}}
 \catcode`Ü=\active
 \defÜ{\"{U}}
 \catcode`Þ=\active
 \defÞ{\c{S}}
 \catcode`þ=\active
 \defþ{\c{s}}
 \catcode`ý=\active
 \defý{{\i}}
 \catcode`ç=\active
 \defç{\d{c}}
 \catcode`Ç=\active
 \defÇ{\d{C}}

\begin{document}

\title{Bound States of the Klein-Gordon Equation for Woods-Saxon
Potential With Position Dependent Mass}

\author{\small Altuð Arda}
\email[E-mail: ]{arda@hacettepe.edu.tr}\affiliation{Department of
Physics Education, Hacettepe University, 06800, Ankara,Turkey}
\author{Ramazan SEVER}
\email[E-mail: ]{sever@metu.edu.tr}\affiliation{Department of
Physics, Middle East Technical University, 06800, Ankara,Turkey}

\date{\today}

\begin{abstract}
The effective mass Klein-Gordon equation in one dimension for the
Woods-Saxon potential is solved by using the Nikiforov-Uvarov
method. Energy eigenvalues and the corresponding eigenfunctions are
computed. Results are also given for the constant mass case.\\
Keywords: Klein-Gordon Equation, Woods-Saxon potential, position
dependent mass, PT-symmetry, energy eigenvalues, eigenfunctions,
Nikiforov-Uvarov method,
\end{abstract}
\pacs{03.65.Fd, 03.65.Ge}

\maketitle

\newpage

\section{Introduction}
The solutions of the non-relativistic and relativistic wave
equations  have attracted attentions in recent years. Furthermore,
the analytical solutions of non-linear equations have also much
attention. For example, the non-linear Landau-Lifshitz equation is
solved to understand the dynamics of Bose-Einstein condansates (BEC)
in an optical lattice [1]. The two-component BEC with attractive
interactions between atoms can be described by a non-linear
Schrödinger equation (SE) called Gross-Pitaevskii equation [2].

Various methods are used to solve the Schrödinger equation based on
perturbative and non-perturbative approaches such as the
hypervirial-Pade summation method [3, 4], group-theoretical approach
[5], Hill determinant method [6], and supersymmetric approaches [7].
The Klein-Gordon (KG) and Dirac equations are also studied for the
Aharonov-Bohm (AB) potential [8], the AB plus the Dirac monopole
potential [9], kink-like [10], Coulomb [11], vector plus scalar
inversely linear potentials [12], PT-symmetric generalized
Wood-Saxon (WS) [13], generalized Hulthen [14], and Rosen-Morse-type
potentials [15]. These solutions are taken in general for constant
mass [16,17]. On the other hand, position dependent mass case has
also many application in different areas, such as impurities in
crystals [18-20], the dependence of nuclear forces on the relative
velocity of the two nucleons [21, 22], or the study of electronics
properties of quantum wells and quantum dots [23], $^3$He clusters
[24], quantum liquids [25] and semiconductor heterostructures [26].
This is also to get the energy eigenvalues and eigenfunctions
[27-31].

Here we intend to solve the KG-equation for the Woods-Saxon
potential in the case of a exponentially mass distribution varying
with position. In nuclear physics, the WS-potential is used to
construct a shell model to describe the single-particle motion in a
fusing system [32].

In the present work, we have obtained the energy spectrum and
corresponding energy eigenfunctions by using the
Nikiforov-Uvarov(NU)-method. We have also obtained the results for
the constant mass case. The NU-method is developed to solve the
second order linear differential equations with special orthogonal
functions. The method is based on solving the equation by reducing
to a generalized equation hypergeometric type [40].

The organization of this work is as follows. In Section II, we solve
the KG-equation in the case of the WS-potential for the spatially
dependent mass by applying the NU-method, and give the energy
eigenvalues and the corresponding eigenfunctions. Our concluding
remarks are given in Section III.

\section{Nikiforov-Uvarov Method and Calculations}
\subsection{Deformed Woods-Saxon Potential}
The KG-equation in one dimension for a particle reads

\begin{eqnarray}
\left[\,\frac{d^2}{dx^2}\,-\,\frac{1}{\hbar^2c^2}\,
[m^2c^4-(E-V)^2]\right]\phi(x)=0\,,
\end{eqnarray}
where $E$ is the energy of the particle, $c$ is the velocity of the
light. The Woods-Saxon potential is

\begin{eqnarray}
V(x)=\,-\,\frac{V_0}{1+qe^{-\beta x}}\,,\,\,\,\,\,(-\infty \leq x
\leq \infty)\,.
\end{eqnarray}
is widely used in the coupled-channels calculations in heavy-ion
physics. This model explains the single-particle motion during a
heavy-ion collisions [32-35]. In this form of the potential, $V_0$
is the potential depth, $q$ is real parameter which determines the
shape of the potential, and $\beta$ is a short notation, i.e.
$\beta\equiv 1/a$, where $a$ is diffuseness of the nuclear surface.

Various mass-distributions are used in the literature. These are
exponential, quadratic [28], inversely-quadratic [36], trigonometric
mass-distributions [37], and mass function of the form
$m(r)=r^{\alpha}$, is especially used for three-dimensional problems
[37, 38]. Here, we prefer to use the following position dependent
mass

\begin{eqnarray}
m(x)=m_0\,\left[1+\,\frac{1}{m_0}\,\left(\,\frac{1+qe^{-\beta
x}}{m_1}\right)^{-1}\,\right]\,,
\end{eqnarray}
This provides us an exact solution of the effective KG-equation.
$m_0$ and $m_1$ in this distribution are two arbitrary positive
parameters. $m_0$ will correspond to the constant mass of the
particle. The mass function is finite at infinity.

Substituting Eqs. (3) and (2) into Eq. (1) we get

\begin{eqnarray}
\frac{d^2\phi(x)}{dx^2}+\left[\,\frac{E^2}{\hbar^2c^2}\,-\,\frac{m_0^2c^4}{\hbar^2c^2}
\,+\,\frac{2EV_0/\hbar^2c^2-2m_0m_1c^4/\hbar^2c^2}{(1+qe^{-\beta x
})}\,+\,\frac{V_0^2/\hbar^2c^2-m_1^2c^4/\hbar^2c^2}{(1+qe^{-\beta x
})^2}\right]\phi(x)=0\,.
\end{eqnarray}
To solve this equation, we use the transformation $z=(1+qe^{-\beta
x})^{-1}$. By defining the following parameters

\begin{eqnarray}
a_3^2=Q^2(m_0^2c^4-E^2)\,,\,\,\,\,\,a_2^2=Q^2(2m_0m_1c^4-2EV_0)\,,\,\,\,\,\,a_1^2=Q^2(m_1^2c^4-V_0^2)\,,
\end{eqnarray}
we obtain

\begin{eqnarray}
\frac{d^2\phi(z)}{dz^2}\,+\,\frac{1-2z}{z-z^2}\,\frac{d\phi(z)}{dz}\,+\,\frac{1}{(z-z^2)^2}
\left[-a_1^2z^2-a_2^2z-a_3^2\right]\phi(z)=0\,.
\end{eqnarray}
where $Q^2=1/\beta^2\hbar^2c^2$. Now to apply the NU-method [40],
we rewrite Eq. (6) in the following form

\begin{eqnarray}
\phi^{\prime\prime}(z)+\,\frac{\tilde{\tau}(z)}{\sigma(z)}\,\phi^{\prime}(z)
+\,\frac{\tilde{\sigma}(z)}{\sigma^2(z)}\,\phi(z)=0,
\end{eqnarray}
where $\sigma(z)$ and $\tilde{\sigma}(z)$ are polynomials with
second-degree, at most, and $\tilde{\tau}(z)$ is a polynomial with
first-degree. We define a transformation for the total wave function
as

\begin{eqnarray}
\phi(z)=\xi(z)\psi(z).
\end{eqnarray}
Thus Eq. (7) is reduced to a hypergeometric type equation

\begin{eqnarray}
\sigma(z)\psi^{\prime\prime}(z)+\tau(z)\psi^{\prime}(z)+\lambda\psi(z)=0.
\end{eqnarray}
We also define the new eigenvalue for the Eq. (7) as

\begin{eqnarray}
\lambda&=&\lambda_n=-n\tau^{\prime}-\,\frac{n(n-1)}{2}\,\sigma^{\prime\prime}\,,
(n=0, 1, 2, \ldots)
\end{eqnarray}
Where

\begin{eqnarray}
\tau(z)&=&\tilde{\tau}(z)+2\pi(z).
\end{eqnarray}
The derivative of $\tau(z)$ must be negative. $\lambda(\lambda_n)$
is obtained from a particular solution of the polynomial $\psi_n(z)$
with the degree of $n$. $\psi_n(z)$ is the hypergeometric type
function whose solutions are given by [40]

\begin{eqnarray}
\psi_n(z)=
\,\frac{b_n}{\rho(z)}\,\frac{d^n}{dy^n}[\sigma^n(z)\rho(z)],
\end{eqnarray}
where the weight function $\rho(z)$ satisfies the equation

\begin{eqnarray}
\frac{d}{dz}[\sigma(z)\rho(z)]=\tau(z)\rho(z).
\end{eqnarray}
On the other hand, the function $\xi(z)$ satisfies the relation

\begin{eqnarray}
\xi^{\prime}(z)/\xi(z)=\pi(z)/\sigma(z).
\end{eqnarray}
Comparing Eq. (6) with Eq. (7), we have
\begin{eqnarray}
\tilde{\tau}(z)=1-2z\,,\,\,\,\,\,\sigma(z)=z(1-z)\,,\,\,\,\,\,
\tilde{\sigma}(z)=-a_1^2z^2-a_2^2z-a_3^2
\end{eqnarray}
It becomes

\begin{eqnarray}
\pi(z)=\,\frac{\sigma^{\prime}(z)-\tilde{\tau}(z)}{2}\,
\pm\,\sqrt{(\frac{\sigma^{\prime}(z)-\tilde{\tau}(z)}{2})^2-\tilde{\sigma}(z)+k\sigma(z)}\,,
\end{eqnarray}
or, explicitly

\begin{eqnarray}
\pi(z)=\mp\sqrt{(a_1^2-k)z^2+(a_2^2+k)z+a_3^2}\,,
\end{eqnarray}
The constant $k$ is determined by imposing a condition such that the
discriminant under the square root should be zero. The roots of $k$
are $k_{1,2}=-a_2^2-2a_3^2\mp2a_3A$, where
$A=\sqrt{a_3^2+a_2^2+a_1^2}$. Substituting these values into
Eq.(16), we get for $\pi(z)$

\begin{eqnarray}
\pi(z)=\,\pm \left \{
\begin{array}{l}(A-a_3)z+a_3,
 \, for \, k_{1}=\,-a_2^2-2a_3^2+2a_3A \\
(A+a_3)z+a_3, \, for \, k_{2}=\,-a_2^2-2a_3^2-2a_3A
\end{array} \right \}.
\end{eqnarray}
Now we calculate the polynomial $\tau(z)$  from $\pi(z)$ such that
its derivative with respect to z must be negative. thus we take the
first choice

\begin{eqnarray}
\tau(z)=1-2(1+A-a_3)z-2a_3.
\end{eqnarray}
The constant $\lambda=k+\pi^{\prime}(z)$ becomes

\begin{eqnarray}
\lambda=-a_2^2+(2a_3-1)(A-a_3)\,,
\end{eqnarray}
and Eq. (10) gives us

\begin{eqnarray}
\lambda_n=2n(1+A-a_3)+n^2-n\,.
\end{eqnarray}
Substituting the values of the parameters given by Eq. (5), and
setting $\lambda=\lambda_n$, one can find the energy eigenvalues
as

\begin{eqnarray}
E_n&=&\,V_0(Q^2m_0m_1c^4-\kappa/2)\tilde{\gamma}\nonumber
\\ &\mp& \, \gamma\tilde{\gamma}\sqrt{4\kappa
m_0m_1c^4-(\kappa^2/Q)^2+4m_0^2(1/\tilde{\gamma}-Q^2m_1^2c^4)}
\end{eqnarray}
where

\begin{eqnarray}
\kappa=\,\frac{1}{2}\,(2n+1)\left[\,\frac{1}{2}\,(2n+1)\mp
\sqrt{1+4a_1^2}\,\right]+\,\frac{1}{4}\,,\,\,\,\,\,\gamma=\sqrt{\kappa+a_1^2}\,,
\,\,\,\,\,\tilde{\gamma}=\,\frac{1}{Q^2V^2_0+\gamma^2}\,.
\end{eqnarray}
We see that the energy levels for particles and antiparticles are
symmetric about
$\frac{V_0(Q^2m_0m_1c^4-\kappa/2)}{Q^2V_0^2+\gamma^2}$. The ground
state energy is different from zero. To have a real energy spectra
we impose

\begin{eqnarray}
(\kappa^2/2Q)^2+Q^2m^2_0m^2_1c^4<\kappa
m_0m_1c^4+m^2_0/\tilde{\gamma}\,,
\end{eqnarray}
We plot four figures to present variation of first three energy
eigenvalues as a functions potential parameters $V_0$ and $\beta$.
Results are agreement with the ones obtained in the literature [39].
$M$ is the ratio $m_1/m_0$, and 'p' and 'a' in the brackets
represent 'particle'and 'antiparticle' in figures.

Now let us find the eigenfunctions. We first compute the weight
function from Eqs. (15) and (19)

\begin{eqnarray}
\rho(z)=z^{-2a_3}(1-z)^{2A}\,,
\end{eqnarray}
and the wave function becomes

\begin{eqnarray}
\psi_n(z)=\,\frac{b_n}{z^{-2a_3}(1-z)^{2A}}\,\frac{d^n}{dz^n}\,\left[
\,z^{n-2a_3}\,(1-z)^{n+2A}\right]\,.
\end{eqnarray}
where $b_n$ is a normalization constant. The polynomial solutions
can be written in terms of the Jacobi polynomials [41]

\begin{eqnarray}
\psi_n(z)=b_n\,P_n^{(2A,\,
-2a_3)}\,(1-2z)\,,\,\,\,\,\,2A>-1\,,\,\,\,\,\,-2a_3>-1\,.
\end{eqnarray}
On the other hand, the other part of the wave function is obtained
from the Eq.(14) as

\begin{eqnarray}
\xi(z)=z^{a_3}\,(1-z)^{A}\,.
\end{eqnarray}
Thus, the total eigenfunctions take

\begin{eqnarray}
\phi_n(z)=b'_n\,(1-z)^{A}z^{a_3}\,P_n^{(2A,\,-2a_3)}\,(1-2z)
\end{eqnarray}
where $b'_n$ is the new normalization constant. It is obtained from

\begin{eqnarray}
\int_{0}^{1}\left|\phi_n(z)\right|^2dz=1\,.
\end{eqnarray}
To evaluate the integral, we use the following representation of the
Jacobi polynomials [42]

\begin{eqnarray}
P_n^{(\sigma,\varsigma)}(z)=\,\frac{\Gamma(n+\sigma+1)}{n!\Gamma(n+\sigma+\varsigma+1)}
\,\sum_{r=0}^{n}\,\frac{\Gamma(n+1)}{\Gamma(r+1)\Gamma(n-r+1)}\,
\frac{\Gamma(n+\sigma+\varsigma+r+1)}{\Gamma(r+\sigma+1)}\,(\,\frac{z-1}{2})^r\,,\nonumber\\
\end{eqnarray}
Hence, from Eq. (30), and with the help of Eq. (31), we get

\begin{eqnarray}
F_{nr}^{2A}\times
F_{ms}^{2A}\left|b'_n\right|^2\int_{0}^{1}z^{2a_3+r+s}\,(1-z)^{2A}\,dz=1\,,
\end{eqnarray}
where $F_{nr}^{2A}$, and $F_{ms}^{2A}$ are two arbitrary functions
of the parameters $A$, and $a_3$, and given by

\begin{eqnarray}
F_{nr}^{2A}=\,\frac{\Gamma(n+2A+1)}{n!\Gamma(n+2A-2a_3+1)}
\,\sum_{r=0}^{n}\,\frac{\Gamma(n+1)}{\Gamma(r+1)\Gamma(n-r+1)}\,
\frac{\Gamma(n+2A-2a_3+r+1)}{\Gamma(r+2A+1)}\,(-1)^r\,,\nonumber\\
\end{eqnarray}
and

\begin{eqnarray}
F_{ms}^{2A} \rightarrow F_{nr}^{2A}\,(n \rightarrow m; r
\rightarrow s)\,.
\end{eqnarray}
The integral in Eq. (32) can be evaluated from the definition of
the Beta function [43]

\begin{eqnarray}
B(\mu,\nu)=\int_0^1y^{\mu-1}\,(1-y)^{\nu-1}\,dy=\,\frac{\Gamma(\mu)\Gamma(\nu)}{\Gamma(\mu+\nu)}
\,,\,\,\,\,\,Re(\mu)>0\,,\,\,\,\,\,Re(\nu)>0\,.
\end{eqnarray}
which gives us

\begin{eqnarray}
\int_0^1z^{2a_3+r+s}\,(1-z)^{2A}\,dz=\,\frac{\Gamma(\tilde{\mu}+r+s)\Gamma(\tilde{\nu})}
{\Gamma(\tilde{\mu}+\tilde{\nu}+r+s)}\,,
\end{eqnarray}
where $\tilde{\mu}=2a_3+1$, and $\tilde{\nu}=2A+1$.

To get the energy eigenvalues for the constant mass case, we set
$q=1$, and $m_1=0$

\begin{eqnarray}
E^{m_1=0}_n=\,-\,\frac{V_0}{2}\mp\,\left[\,\sqrt{\beta^2-4V^2_0}-\beta(2n+1)\right]
\left[\,\frac{m^2_0}{4V_0^2+[\sqrt{\beta^2-4V^2_0}-\beta(2n+1)]^2}\,-\,\frac{1}{16}\right]^{1/2}\,.
\end{eqnarray}
This is the same  in Eq. (46) in Ref. [14].

Since the wave function changes only with the parameter $A$, we
simply get

\begin{eqnarray}
\phi^{m_1=0}_n(z)=b''_n\,(1-z)^{A'}\,z^{a_3}\,P_n^{(2A',\,-2a_3)}(1-2z)\,,
\end{eqnarray}
where the new parameter
$A'=\sqrt{a_3^2-2\varrho^2E/V_0-\varrho^2}$.

\subsection{Non-$PT$ Symmetric and non-Hermitian deformed Woods-Saxon Potential }
In this case, we take the potential parameters as $V_0 \rightarrow
iV_0$, and $\beta \rightarrow \beta$. So, the potential takes the
form [44]

\begin{eqnarray}
V(x)=\,-\,\frac{iV_0}{1+qe^{-\beta x}}\,,
\end{eqnarray}
From Eq. (3), we obtain

\begin{eqnarray}
\frac{d^2\phi(x)}{dx^2}+\left[\,\frac{E^2}{\hbar^2c^2}\,-\,\frac{m_0^2c^4}{\hbar^2c^2}
\,+\,\frac{2iEV_0/\hbar^2c^2-2m_0m_1c^4/\hbar^2c^2}{(1+qe^{-\beta
x
})}\,-\,\frac{V_0^2/\hbar^2c^2+m_1^2c^4/\hbar^2c^2}{(1+qe^{-\beta
x })^2}\right]\phi(x)=0\,.\nonumber\\
\end{eqnarray}
By using the same coordinate transformation and defining the
following parameters

\begin{eqnarray}
-a_3^2=Q^2(E^2-m_0^2c^4)\,,\,\,\,\,\,-A_2^2=Q^2(2iEV_0-2m_0m_1c^4)\,,\,\,\,\,\,-A_1^2=Q^2(-m_1^2c^4-V_0^2)\,,
\end{eqnarray}
we get

\begin{eqnarray}
\frac{d^2\phi(z)}{dz^2}\,+\,\frac{1-2z}{z-z^2}\,\frac{d\phi(z)}{dz}\,+\,\frac{1}{(z-z^2)^2}
\left[-A_1^2z^2-A_2^2z-a_3^2\right]\phi(z)=0\,.
\end{eqnarray}

Following the same procedure, we find the energy spectra

\begin{eqnarray}
E_n&=&\,\frac{iV_0(2m_0m_1c^4-\kappa''Q^2)}{2\zeta}
\nonumber \\
&\pm&\,\frac{1}{2\zeta}\,\sqrt{V_0^2(2m_0m_1-\kappa''Q^2)^2-
\zeta(\kappa''^2-4\kappa''Q^2m_0m_1c^4+4m_0^2c^4Q^2(Q^2m_1^2-\kappa''-A_1^2))}\,,\nonumber\\
\end{eqnarray}
where

\begin{eqnarray}
\zeta=Q^2(\kappa''+Q^2m_1^2c^4)\,\,\,\,\,,\kappa''=\,\frac{1}{2}\,(2n+1)\left[\,\frac{1}{2}\,(2n+1)
+\sqrt{1+4A_1^2}\,\right]\,+\,\frac{1}{4}\,.
\end{eqnarray}
and the corresponding total wave functions as

\begin{eqnarray}
\phi_n(z)=b'''_n\,(1-z)^B\,z^{a_3}\,P_n^{(2B,\,-2a_3)}\,(1-2z)\,,
\end{eqnarray}
where $B=\sqrt{A_1^2+A_2^2+a_3^2}$. It is seen that the energy
eigenvalues are consist of the real and imaginary parts. For the
constant mass case, we have

\begin{eqnarray}
E^{m_1=0}_n=\,-\,\frac{iV_0}{2}\,\pm\,\frac{1}{2Q^2\kappa''}\,\sqrt{\kappa''^2Q^2(\varrho^2-\kappa'')
+4m_0^2c^4Q^4\kappa''(\kappa''+\varrho^2)}\,,
\end{eqnarray}
and the total eigenfunctions are

\begin{eqnarray}
\phi^{m_1=0}_n(z)=b''''_n\,(1-z)^{B'}\,z^{a_3}\,P_n^{(2B',\,-2a_3)}\,(1-2z)\,,
\end{eqnarray}
where $B'=\sqrt{a_3^2-2iE\rho^2/V_0+\rho^2}$. The energy spectra
have real and imaginary part in the constant mass case. The
imaginary part does not depend the quantum number $n$. The
normalization constant $b'''_n$ is also obtained in the same way.
That is

\begin{eqnarray}
G_{nr}^{2B} \times
G_{ms}^{2B}\,|b'''_n|^2\,\int_0^{1}\,z^{2a_3+r+s}\,(1-z)^{2B}\,dz=1\,,
\end{eqnarray}
The integral can be evaluated by using Eq. (35)

\begin{eqnarray}
\int_0^{1}\,z^{2a_3+r+s}\,(1-z)^{2B}\,dz=\,\frac{\Gamma(m'+r+s)\Gamma(n')}
{\Gamma(m'+n'+r+s)}\,,
\end{eqnarray}
where $m'=2a_3+1$, and $n'=2B+1$. Two functions $G_{nr}^{2B}$, and
$G_{ms}^{2B}$ are given by

\begin{eqnarray}
G_{nr}^{2B}&\rightarrow F_{nr}^{2A}(2A \rightarrow 2B)\,,\nonumber
\\ G_{ms}^{2B}&\rightarrow F_{ms}^{2A}(2A \rightarrow 2B)\,.
\end{eqnarray}
where the functions $F_{nr}^{2A}$, and $F_{ms}^{2A}$ are defined in
Eqs. (33) and (34).

\section{Conclusion}
We have solved the one dimensional effective mass KG-equation for
the Woods-Saxon potential. The energy spectra and the corresponding
wave functions are obtained by applying the NU-method. We have found
a real energy spectra for the WS-potential in the position dependent
mass case. To check our results, we have also calculated the energy
eigenvalues of the particle and antiparticles for the constant mass
limit. We have also studied the non-$PT$ symmetric and non-Hermitian
case. We have seen that the energy spectra have real and imaginary
parts in this case. We have also obtained the energy spectra and
corresponding eigenfunctions for the constant mass limit for this
case.

\section{Acknowledgments}
This research was partially supported by the Scientific and
Technical Research Council of Turkey.

\newpage

\begin{figure}[htbp]
\centering
\includegraphics[height=5in, width=7in]{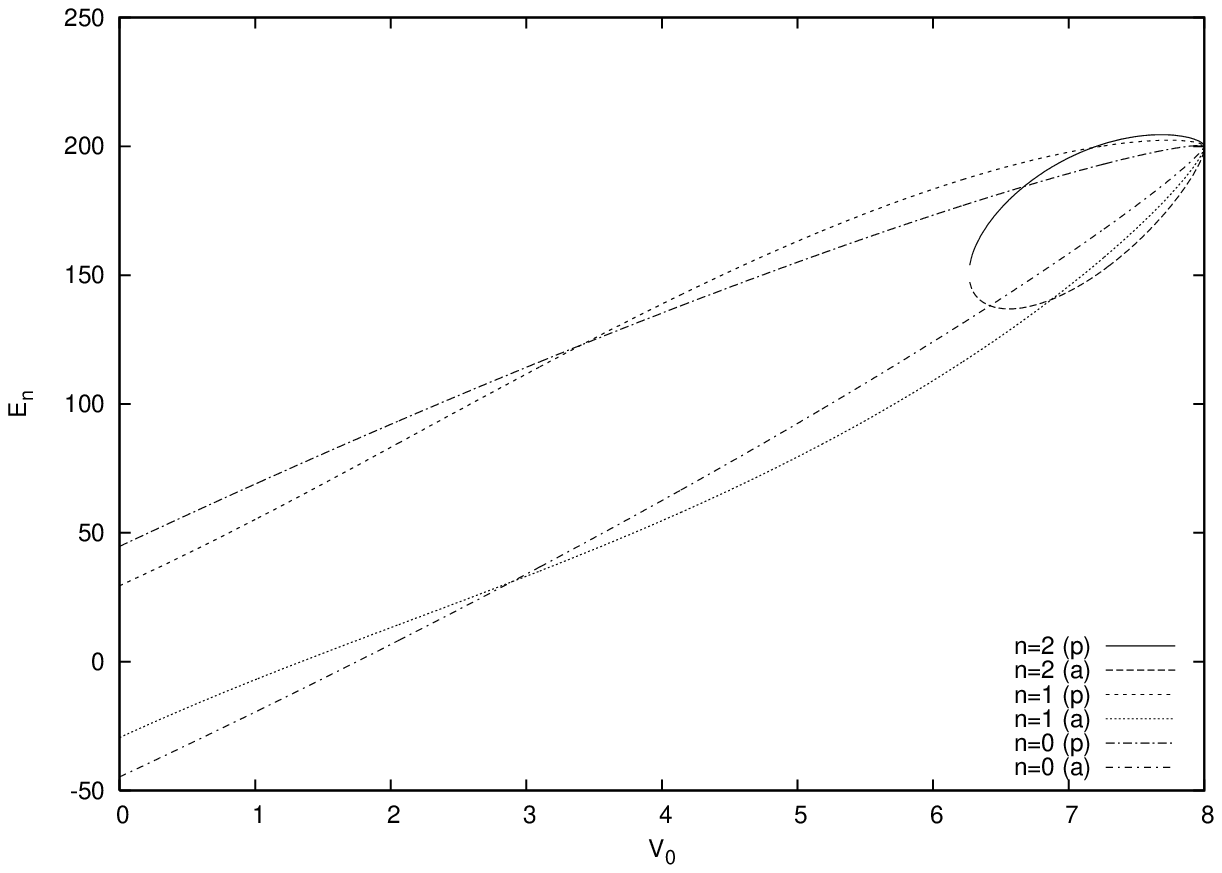}
\caption{The dependence of first three excited states on $V_0$ in
the case of $M=0.04$, and $\beta=0.1$.}
\end{figure}

\newpage

\begin{figure}[htbp]
\centering
\includegraphics[height=5in, width=7in]{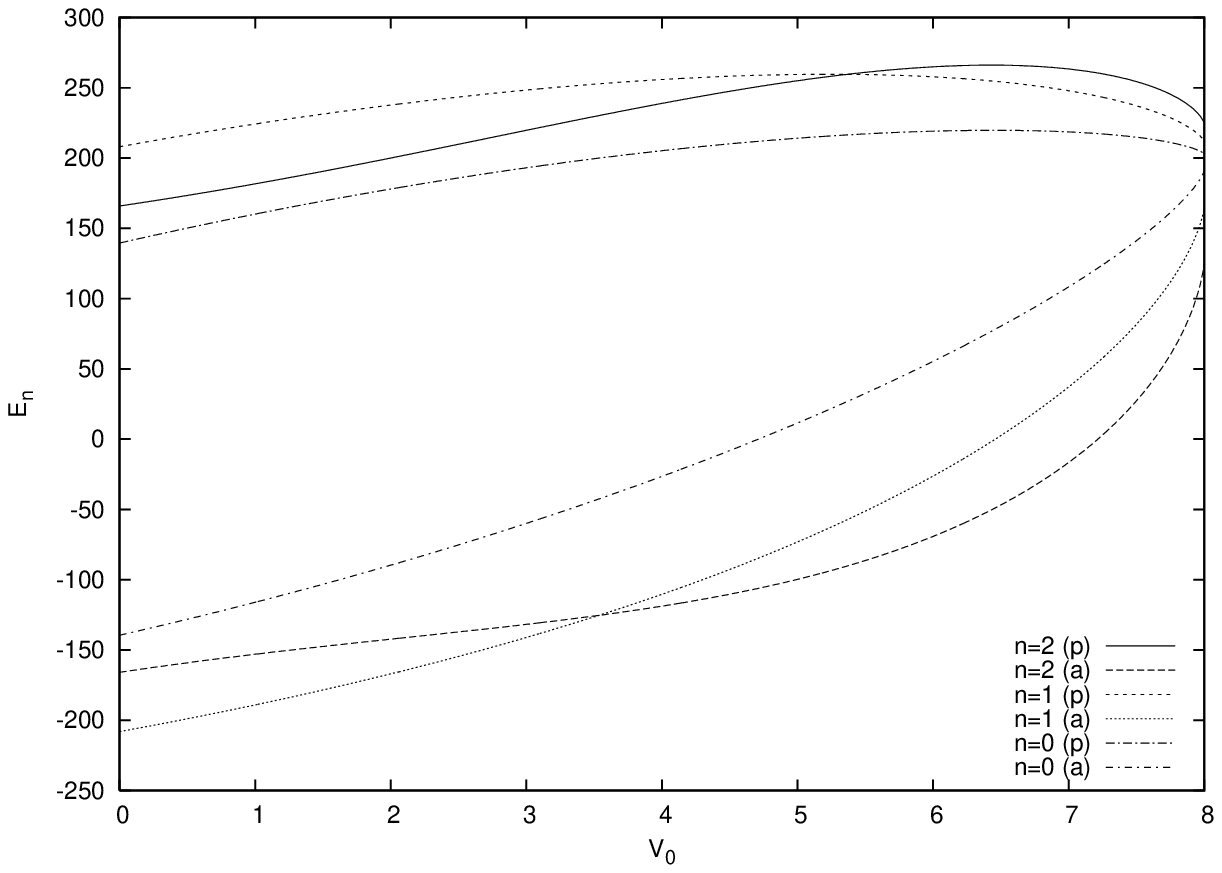}
\caption{The dependence of first three excited states on $V_0$ in
the case of $M=0.04$, and $\beta=1$.}
\end{figure}

\newpage

\begin{figure}[htbp]
\centering
\includegraphics[height=5in, width=7in]{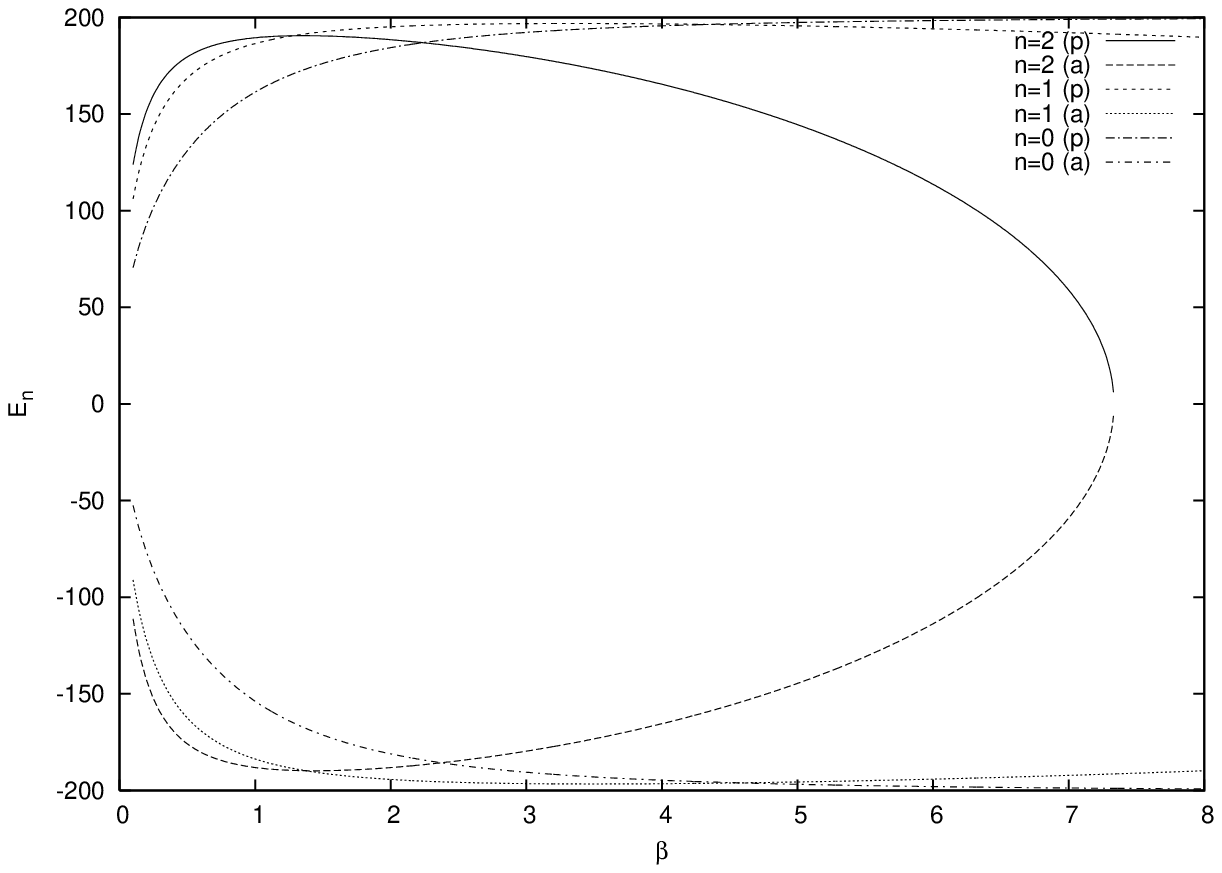}
\caption{The dependence of first three excited states on $\beta$
in the case of $M=0.01$, and $V_0=0.1$.}
\end{figure}

\newpage

\begin{figure}[htbp]
\centering
\includegraphics[height=5in, width=7in]{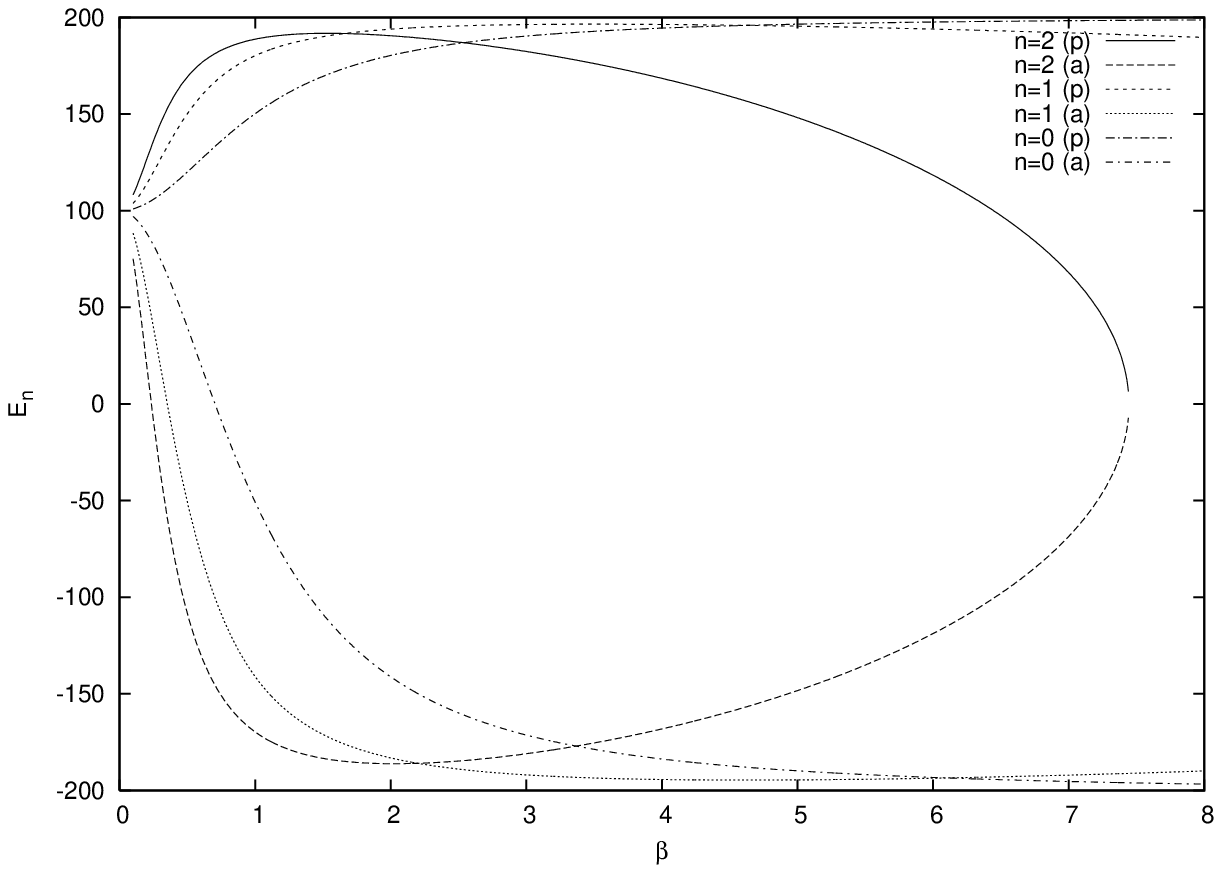}
\caption{The dependence of first three excited states on $\beta$
in the case of $M=0.01$, and $V_0=1$.}
\end{figure}


\newpage

\begin{thebibliography}{99}

\bibitem{ref1} Z.~D.~Li, P.~B.~He, L.~Li, J.~Q.~Liang and W.~M.~Liu, Phys. Rev. A {\bf 71}, 053611
(2005).

\bibitem{ref2} L.~Li, B.~A.~Malomed, D.~Mihalache and W.~M.~Liu, Phys. Rev. E {\bf 73}, 066610
(2006).

\bibitem{ref3} J.~P.~Killingbeck, A.~Grosjean, and G.~Jolicard,
J. Phys. A: Math. Gen. {\bf 34}, 8309 (2001).

\bibitem{ref4} A.~Arda, Turk. J. Phys. {\bf 28}, 223 (2004).

\bibitem{ref5} J.~Chen, L.~C.~Kweck and C.~H.~Oh, Phys. Rev. A {\bf 67}, 012101
(2003).

\bibitem{ref6} S.~N.~Biswas et al., J. Math. Phys. {\bf 14}, 1190 (1973).

\bibitem{ref7} A.~Khare, U.~P.~Sukhatme, J. Phys. A {\bf 26}, L901-L904 (1993)
[arXiv: hep-th/9212147].

\bibitem{ref8} C.~R.~Hagen, Phys. Rev. D {\bf 48}, 5935 (1993),
[arXiv:~hep-th/9308009].

\bibitem{ref9} V.~M.~Villalba, [arXiv:~hep-th/9409102].

\bibitem{ref10} A.~S~.~de~Castro, [arXiv:~hep-th/0511010] and references therein.

\bibitem{ref11} A.~S~.~de~Castro, Phys. Lett. A {\bf 346}, 71 (2005),
[arXiv:~hep-th/0507218].

\bibitem{ref12} A.~S~.~de~Castro, Phys. Lett. A {\bf 338}, 81 (2005),
[arXiv:~hep-th/0502201].

\bibitem{ref13} S.~M.~Ikhdair and R.~Sever, Ann. Phys. {\bf 16}, 218 (2007).

\bibitem{ref14} M.~Þimþek and H.~Eðrifes, J. Phys. A {\bf 37}, 4379 (2004).

\bibitem{ref15} L.-Z.~Yi, Y.-F.~Diao,J.-Y.~Liu, and C.-S.~Jia, Phys. Lett. A {\bf
333}, 212 (2004).

\bibitem{ref16} A.~Mostafazadeh, J. Phys. A {\bf 31}, 6495 (1998).

\bibitem{ref17} A.~Mostafazadeh, Phys. Rev. A {\bf 55}, 4084 (1997).

\bibitem{ref18} J.~M.~Luttinger and W.~Kohn, Phys. Rev. {\bf 97}, 869 (1955).

\bibitem{ref19} G.~H.~Wanner, Phys. Rev. {\bf 52}, 191 (1957).

\bibitem{ref20} J.~C.~Slater, Phys. Rev. {\bf 76}, 1592 (1949).

\bibitem{ref21} O.~Rojo and J.~S.~Levinger, Phys. Rev. {\bf 123}, 2177 (1961).

\bibitem{ref22} M.~Razavy, G.~Field, and J.~S.~Levinger, Phys. Rev. {\bf 125}, 269
(1962).

\bibitem{ref23} L.~Serra and E.~Lipparini, Europhys. Lett. {\bf 40}, 667 (1997).

\bibitem{ref24} M.~Barranco et al., Phys. Rev. B{\bf 56}, 8997 (1997).

\bibitem{ref25} F.~Arias~de~Saavedra et al., Phys. Rev. B {\bf 50}, 4248 (1994).

\bibitem{ref26} T.~Gora and F.~Williams, Phys. Rev. {\bf 177}, 11979 (1969);
O.~Von~Roos, Phys. Rev. B {\bf 27}, 7547 (1983).

\bibitem{ref27} A.~D.~Alhaidari, Phys. Lett. A {\bf 322}, 72 (2004).

\bibitem{ref28} A.S.~Dutra and C.~A.~S.~Almeida, Phys. Lett. A {\bf 275}, 25
(2000).

\bibitem{ref29} B.~Gonul, B.~Gonul, D.~Tutcu, and O.~Ozer, Mod. Phys. Lett. A {\bf
17}, 2057 (2002); B.~Gonul, O.~Ozer, B.~Gonul, and F.~Uzgun, Mod.
Phys. Lett. A {\bf 17}, 2453 (2002); B.~Gonul and M.~Kocak, Chin.
Phys. Lett. {\bf 20}, 2742 (2005); B.~Gonul and M.~Kocak,
arXiv:~quant-ph/0512035.

\bibitem{ref30} C.~Tezcan and R.~Sever, [arXiv:~quant-ph/0604041].

\bibitem{ref31} I.~O.~Vakarchuk, J. Phys. A: Math. Gen. {\bf 38}, 4727 (2005).

\bibitem{ref32} A.~Diaz-Torres, and W.~Scheid, Nucl. Phys. A {\bf 757}, 373
(2005), [arXiv:~nucl-th/0504002].

\bibitem{ref33} C.~Berkdemir, A.~Berkdemir and R.~Sever, Phys. Rev. C {\bf 72}, 027001 (2005).

\bibitem{ref34} K.~Hagino et al., [arXiv:~nucl-th/0110065].

\bibitem{ref35} S.~M.~Ikhdair and R.~Sever, [arXiv:~quant-ph/0507272].

\bibitem{ref36} L.~Jiang, L.-Z.~Yi, and C.-S.~Jua, Phys. Lett. A {\bf 345}, 249
(2005).

\bibitem{ref37} A.~D.~Alhaidari, Phys. Rev. A {\bf 66}, 042116 (2002).

\bibitem{ref38} G.-X.~Ju, Y.~Xiang, and Z.-Z.~Ren, [arXiv: quant-ph/0601005].

\bibitem{ref39} S.~M.~Ikhdair and R.~Sever, [arXiv:~quant-ph/0610183].

\bibitem{ref40} A.~F.~Nikiforov, and V.~B.~Uvarov, \textit{Special Functions of
Mathematical Physics }, (Birkh\"{a}user, Basel, 1988).

\bibitem{ref41} C.~W.~Wong, \textit{Introduction to Mathematical Physics-Methods
and Concepts }, (Oxford University Press, 1991).

\bibitem{ref42} G.~Szegö, \textit{Orthogonal Polynomials }, (Providence, RI: Amer.
Math. Soc., 1988).

\bibitem{ref43} M.~Spiegel, \textit{Theory and Problems of Laplace Transforms },
(New York, 1965).

\bibitem{ref44} A.~Berkdemir, C.~Berkdemir, and R.~Sever, Mod. Phys. Lett. A {\bf
21}, 2087 (2006).
\end{thebibliography}
\end{document}